\documentclass[superscriptaddress, aps, prx, longbibliography]{revtex4-2}

\usepackage{graphicx}
\usepackage{physics}
\usepackage{amsmath}
\usepackage{array}
\usepackage{siunitx}
\usepackage[section]{placeins}
\usepackage{hyperref}
\usepackage{soul}
\usepackage{ulem}
\usepackage{cancel}
\usepackage{afterpage}

\usepackage{natbib}

\begin{document}

\title{Superconducting flux concentrator coils for levitation of particles in the Meissner state}

\date{June 2025}

\author{Robert Smit}
\author{Martijn Janse}
\author{Eli van der Bent}
\author{Thijmen de Jong}
\author{Kier Heeck}
\author{Jaimy Plugge}
\author{Tjerk Oosterkamp} 
\author{Bas Hensen} \email{hensen@physics.leidenuniv.nl}
\address{Leiden Institute of Physics, Leiden University, P.O. Box 9504, 2300 RA Leiden, The Netherlands}

\begin{abstract}
Magnetic levitation of superconductors is a promising platform to study quantum mechanics in the large-mass limit. One major limitation is the weak trapping potential, which results in low vibrational eigenfrequencies and increased susceptibility to low-frequency noise. While generating strong magnetic fields is relatively straightforward, creating a tightly confined harmonic potential — essentially achieving a large magnetic field gradient — remains a significant challenge. In this work, we demonstrate a potential solution using superconducting cores that concentrate magnetic flux into arbitrarily small volumes. We show the ability to trap superconducting particles using an anti-Helmholtz coil configuration incorporating these cores. However, we observe rapid damping of the levitated particle motion due to flux trapping within the cores, occurring once the lower critical field is exceeded locally. To investigate this mechanism, we employ diamond NV center magnetometry and detect substantial remanent fields persisting after high-current operation of the coils. Finally, we discuss possible strategies to mitigate this effect and improve the levitation properties.
\end{abstract}

\maketitle

\section{Introduction}

The levitation of particles is a powerful and widely employed technique for isolating objects from their environment, thereby minimizing dissipative interactions\cite{gonzalez-ballestero_levitodynamics_2021}. The achievement of a long-lived and large-mass motional quantum state would be a milestone towards addressing fundamental questions in physics, including those concerning wave function collapse models\cite{vinante_testing_2019} and the interplay between quantum mechanics with gravity\cite{bose_spin_2017, bose_2025}. Beyond theoretical implications, levitated particles also serve as highly sensitive platforms for detecting gravitational forces\cite{fuchs_measuring_2024}, torques\cite{ahn_ultrasensitive_2020}, and even individual nuclear decay events\cite{wang_mechanical_2024}.\\
\\
The current state-of-the-art in particle levitation is the optical tweezer, which has realized significant milestones over the past decade\cite{ashkin_acceleration_1970, chu_experimental_1986}. A particular achievement is the cooling of the motion of a levitating silica nanoparticle to the quantum ground state, first along a single vibrational mode\cite{delic_cooling_2020, tebbenjohanns_quantum_2021} and later in two modes simultaneously\cite{piotrowski_simultaneous_2023}. Despite these achievements, optical traps inherently suffer from heating via photon absorption\cite{millen_nanoscale_2014} and scattering\cite{jain_direct_2016}, a limitation that becomes increasingly severe for larger, more massive particles, particularly for those approaching the 1 $\mu$m scale, where gravitational effects are expected to become relevant\cite{bose_spin_2017}. \\ 
\\
Alternative levitation techniques, based on electric and magnetic fields, offer potential pathways to circumvent these limitations. According to Earnshaw's theorem, stable levitation of charged or magnetic objects using static fields alone is impossible, necessitating dynamic potentials such as the rotating saddle potential in electric\cite{paul_electromagnetic_1990} and magnetic Paul traps\cite{perdriat_planar_2023, janse_characterization_2024}. However, static magnetic levitation is achievable for diamagnetic particles that acquire an induced dipole in an applied magnetic field\cite{simon_diamagnetic_2000,hsu_cooling_2016}.\\ 
\\
Superconducting particles represent an ideal diamagnetic system, perfectly repelling external magnetic fields by the Meissner effect\cite{hofer_analytic_2019}. Magnetic confinement of such particles has been demonstrated using anti-Helmholtz coil configurations with both superconducting wire loops\cite{hofer_high-q_2023} and planar on-chip coils\cite{gutierrez_latorre_superconducting_2023}. However, the achievable eigenfrequencies of motion within a magnetic trap remain significantly lower than those possible in optical traps. A clear route to increasing them involves higher coil currents, but this approach introduces substantial technical difficulties when operating in a millikelvin environment, more so when including vibration isolation systems \cite{hofer_high-q_2023}. In this work, we explore an alternative strategy: we employ superconducting flux concentrators to locally enhance the magnetic field gradient, thereby enabling stable levitation of superconducting particles without resorting to impractically large currents. Furthermore, we investigate potential dissipation mechanisms by incorporating nitrogen-vacancy (NV) center defects as nanoscale magnetic field sensors. Through this study we provide both a promising route towards enhanced magnetic levitation systems and a detailed characterization of the associated technical challenges. 

\section{Results}

\subsection{The flux concentrator trap}

\begin{figure}
    \centering
    \includegraphics[width=\textwidth]{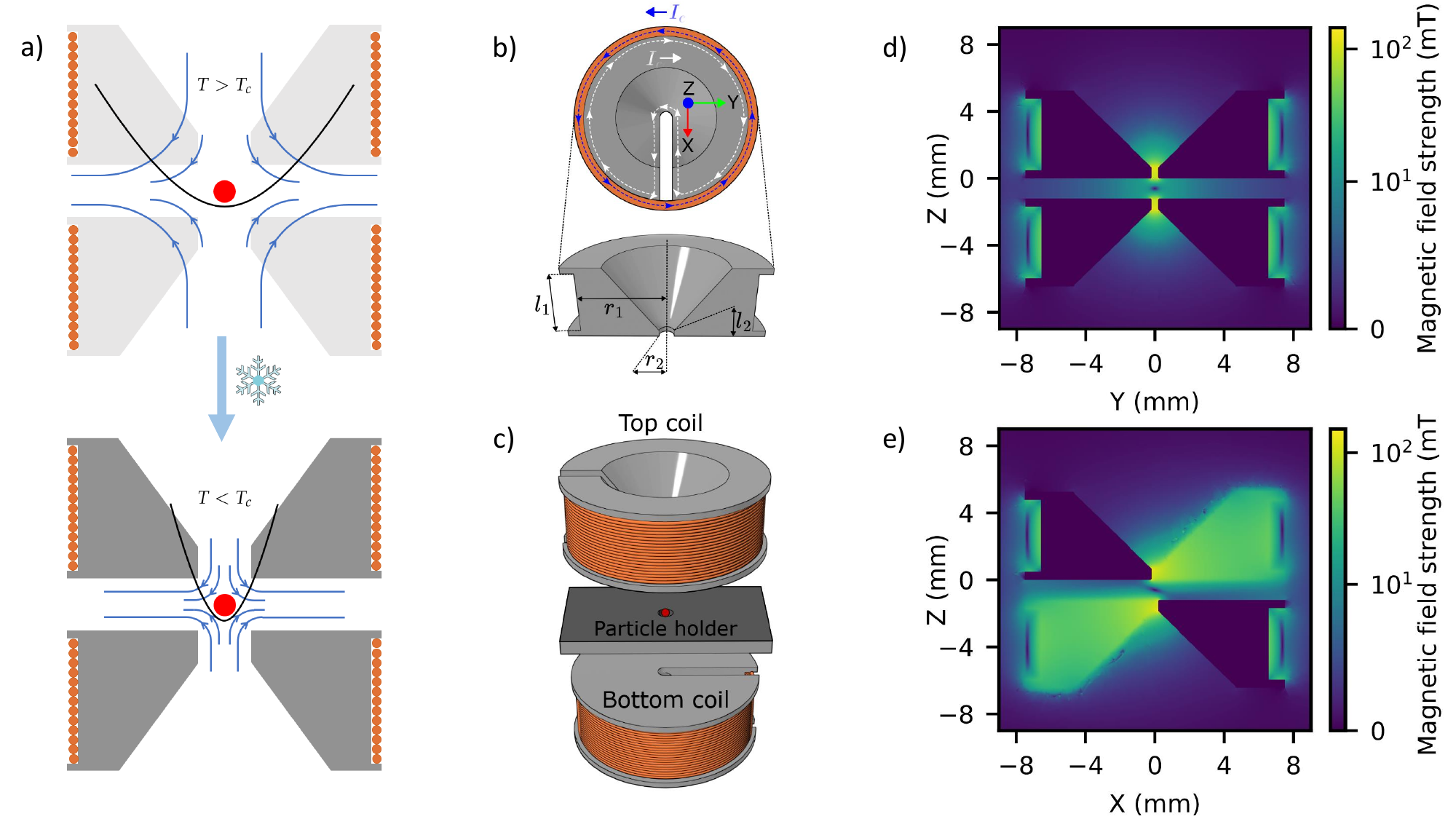}
    \caption{\textbf{Flux concentration to enhance trap stiffness for Meissner levitation in anti-Helmholtz configuration}. \textbf{(a)} Two coils (orange) create a quadrupole magnetic field (blue lines) resulting in a harmonic confinement potential (black line) for a superconducting particle (red dot). When the core enters its superconducting state from $T<T_\text{c}$ onward, the magnetic field flux lines of the coil are concentrated in the slit due to the Meissner effect, enhancing the magnetic field gradient and thus the stiffness of the trapping potential. \textbf{(b)} Equivalently, a current ($I_{\text{coil}}$) through the outer wound coil induces an equal shielding current that revolves around an inner loop, leading to local concentration of the magnetic field. The dimensions of the cores in the experiments are $l_1$ = 4.5 mm, $l_2$ = 0.45 mm, $r_1$ = 6.57 mm and $r_2$ = 0.2 mm. \textbf{(c)} A set of these identical coils, placed in anti-Helmholtz configuration. The spacer is a PEEK plate with the superconducting particle inside. \textbf{(d)-(e)} The simulated potential minimum of the magnetic field for projections on the YZ and XZ axis, according to the axes system shown in panel (b). The simulations correspond to an applied current of 1 A through both coils.}
    \label{Figure1}
\end{figure}

The earliest demonstrations of magnetic flux focusing or concentration involved coils wrapped around electrically conductive materials. These flux concentrators were originally developed to produce high transient magnetic fields\cite{furth_production_1957} or to generate strong localized fields in a confined region\cite{brechna_150_1965}. In our flux concentrator configuration, illustrated in Figure \ref{Figure1}a, the magnetic field produced by an outer coil induces circulating eddy currents within the core. By introducing a slit into the core, extending to its center, the induced current is forced to navigate around a narrow central path, resulting in a locally enhanced magnetic field. The absence of resistive losses and the perfect diamagnetic shielding inherent to superconductors make them highly efficient flux concentrators, a capability demonstrated using a range of superconducting materials\cite{kiyoshi_magnetic_2009, zhang_magnetic_2011}. While flux concentrators can dramatically increase local field strengths, the enhancement persists only until the applied field approaches the critical current of the superconducting core. For levitation applications, however, the field strength itself is less critical than the achievable magnetic field gradient. Notably, superconducting flux concentrators have recently been proposed as a means of generating large field gradients suitable for diamagnetic levitation\cite{takahashi_simulation_2020}. Beyond generating concentrated fields, similar designs have been applied to efficiently collect magnetic flux from localized areas, particularly in scanning magnetic susceptibility measurements, where they are coupled to SQUID sensors\cite{xiang_flux_2023}. \\

In this work, we employ machined niobium cores with dimensions listed in the caption of Figure \ref{Figure1}. The system can be effectively modeled as two separate coils: the outer drive coil and the inner induced current loop within the superconducting core. Assuming that the formula for idealized long solenoids applies, the amplification factor of the magnetic field maximum at the center can be approximated as $\alpha$ = $l_2/l_1$ where $l_2$ and $l_1$ are the characteristic dimensions of the core. Based on our core geometry, we estimate an amplification factor of approximately 10, a value corroborated by finite element simulations (see Supplementary Figure S8a).\\ 
\\
To create a trapping potential suitable for levitating a superconducting particle, a pair of these flux concentrator coils are assembled in an anti-Helmholtz configuration, with their slits anti-aligned and stacked (Figure \ref{Figure1}b). A 1 mm thick PEEK spacer, sandwiched between two 100 $\mu$m cover glasses, separates the coils, leaving a cavity to hold a 50 $\mu$m diameter tin-lead (63:37) alloy particle (EasySpheres), which is expected to transition to a superconducting state below 7 K \cite{livingston_magnetic_1963}. The coils around the niobium cores comprise 180 turns of 100 $\mu$m niobium-titanium wire with copper cladding (SuperCon). \\
\\
Finite element simulations of this double-concentrator anti-Helmholtz configuration confirm the presence of a magnetic field minimum between the two anti-aligned slits (Figure \ref{Figure1}c,\ref{Figure1}d). The modeled inter-coil spacing of 1.2 mm corresponds to the physical thickness of the PEEK spacer and cover glasses. Under these conditions, the calculated field gradients are approximately 22, 34, and 56 T/m per ampere of current along the X, Y and Z axes respectively (see axes definitions in Figure \ref{Figure1}a), values comparable to those achieved with conventional superconducting anti-Helmholtz traps, both with wound wire coils and on-chip planar configurations \cite{gutierrez_latorre_superconducting_2023,hofer_high-q_2023}. \\
\\
For the levitation experiments, the assembled flux concentrator trap was mounted in a copper holder and attached to the millikelvin stage of a dilution refrigerator (see Supplementary Figure S1). An optical access window allowed imaging of the XY plane of the trap using a camera, while illumination was provided from below via an optical fiber. This uniform illumination scheme minimized radiative heating of the levitated particle, enabling it to remain in the superconducting state for extended periods, up to tens of minutes. Superconducting wiring connected the coils to external current supplies, with thermal anchoring at intermediate refrigerator stages. Continuous currents up to 1.7 A could be applied, with levitation typically achieved at currents between 600–800 mA. Initial ring-down measurements following levitation exhibited dominant motion along the slit direction, along which the trapping potential is weakest. Subsequent adjustments to individual coil currents produced shifts in the particle’s equilibrium position along this direction (see Supplementary Figure S2), as expected due to the inclined trapping potential as observed from the simulations projected in the XZ-plane (Figure \ref{Figure1}c). The inclination of the potential can be avoided by matching the vertical separation of the flux concentrator coils to the radii of the coils (see Supplementary Figure S9).\\

\begin{figure}
    \centering
    \includegraphics[scale=1]{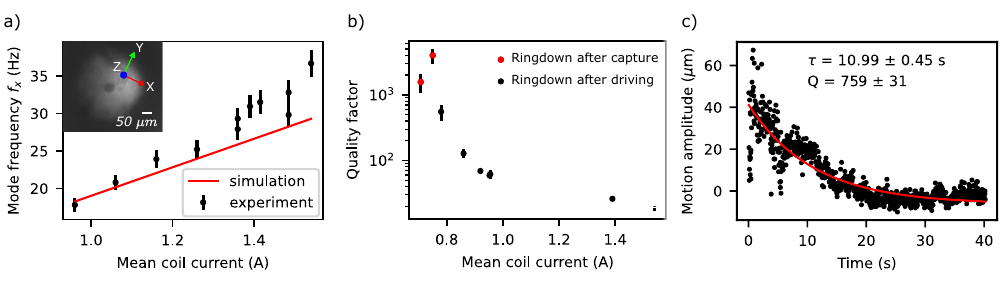}
    \caption{\textbf{Characteristics of levitation}. \textbf{(a)} The dependence of the x-mode (in the x-direction, i.e. along the slit: see axes system in the inset picture of the levitating particle) frequency as a function of the mean current through the coils. The resonance frequency was obtained by tuning the frequency of a driving current through the bottom coil, while optically monitoring the motional amplitude with a camera. \textbf{(b)} A set of measurements of the quality factor of the x-mode for different coil currents. The red points correspond to quality factors obtained from the ringdown after trapping - thus constituting mixed modes - and the black points after driving along the slit direction. \textbf{(c)} An example ringdown, where only the motional amplitude is included. Outlying points are explained by the difficulty of the video-tracking algorithm to track the particle's position when the motion is relatively large. The mean coil current was approximately 0.8 A.}
    \label{Figure2}
\end{figure}

The particle’s vibrational dynamics were further characterized by feeding the bottom coil of the trap with an AC current, supplied via a function generator and a 400 $\Omega$ current-limiting resistor, and tracking the particle motion by video analysis (see Figure S3 in the Supplementary Information for details). The primary resonance mode identified corresponded to motion along the slit direction. Other resonances were either weakly coupled, difficult to resolve optically, or resulted in the particle escaping the trap upon excitation. Consequently, our analysis focused on the x-mode (along the slit). \\
\\
In diamagnetic levitation systems, the vibrational mode frequencies $f_i$ depend on the field gradient in the direction of motion\cite{gutierrez_latorre_superconducting_2023}: $f_i = \nabla_i B \sqrt{3/8\pi \mu_0 \rho} =  \zeta_i \mu_0I/r_2^2 \sqrt{3/8\pi \mu_0 \rho}$. Here, the particle density is given by $\rho$, $r_2$ is the inner radius of the flux concentrator core, and $\zeta_i$ is a geometric factor. The equation reveals that a linear dependence with current is expected for the eigenmode frequencies. With finite element simulations we indeed find a linear relationship with current (see Figure S10 in Supplementary). The simulated eigenmode frequencies result in geometric factors of $\zeta_x$ = 0.18, $\zeta_y$ = 0.36 and $\zeta_z$ = 0.53 and reveal that the trap geometry is significantly anisotropic \cite{hofer_analytic_2019}. The experimental measurements of the x-mode frequencies (shown in Figure \ref{Figure2}a) show an approximately linear relationship with mean coil current. However, we note that the individual measurements were not performed at the same levitation height: the ratio between the separate currents on the two coils is not constant. This may explain the offset of data points from the simulated curve at higher mean coil currents. \\
\\
Interestingly, the measured resonance frequencies and damping rates were not always reproducible, varying by up to 10\% depending on the preceding history of currents applied to the coils (Figure \ref{Figure2}b). Additionally, higher coil currents tended to degrade the system’s quality factor. The most stable and highest quality factors, reaching up to a thousand, were observed in a freshly cooled flux concentrator that was subsequently operated at low currents only (example given in Figure \ref{Figure2}c). Once higher currents were applied, the quality factors declined sharply and remained low until the entire system was thermally cycled, either by heating up the whole cryostat or heating the flux concentrator locally with a fiber-coupled laser. These findings suggest a current-dependent operational regime for the flux concentrators, beyond which an additional or enhanced damping mechanism emerges. The likely cause involves flux penetration into the superconducting niobium cores, especially at geometric boundaries, such as the sharp edges of the slit. Simulations indicate that local magnetic fields can exceed 350 mT at the maximum applied current of 1.7 A, which is significantly higher than the lower critical field $B_{c,1}$ for high-quality niobium, reported at 173.5 mT\cite{stromberg_superconducting_1965}. Given this, flux entry into the superconducting cores is probable at elevated currents, leading to increased dissipation and reduced mechanical quality factors.

\afterpage{
\clearpage
\begin{figure}
    \centering
    \includegraphics[scale=1]{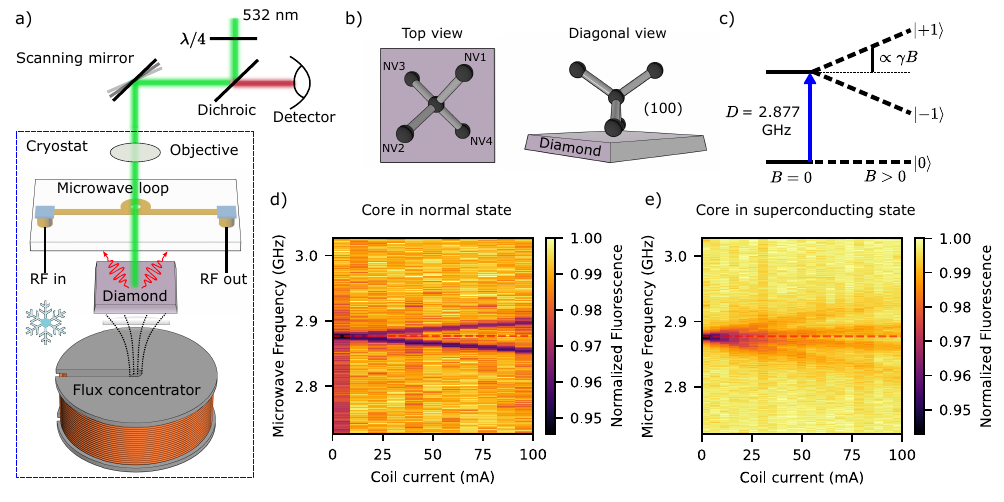}
    \caption{\textbf{Magnetometry with NV centers}. \textbf{(a)} A schematic of the measurement setup and configuration. \textbf{(b)} The bond orientations of the lattice of a (100) cut diamond. \textbf{(c)} The projection of a magnetic field on the NV axis will split the two spin states with $m_s$ = $\pm$1, while their relative frequency with respect to the $m_s$ = 0 state can be read out as an ODMR signal. \textbf{(d)-(e)} Such ODMR spectra plotted as a function of coil current for a core in normal state (d) and in superconducting state (e), where a dashed line represents the center frequency determined by the zero-field splitting $D$ at 2.877 GHz.}
    \label{Figure3}
\end{figure}
}

\subsection{NV center magnetometry}
To investigate local magnetic field distributions and potential flux trapping within the flux concentrators, we employed nitrogen-vacancy (NV) center defects in diamond as nanoscale magnetic field probes. These sensors allow for spatially resolved, non-invasive measurements of magnetic fields within the trapping region. The NV center in diamond is a point defect consisting of a nitrogen atom substituting for a carbon atom, adjacent to a vacant lattice site. In its negatively charged form, denoted NV$^{-}$, the defect possesses a triplet ground electronic state with three spin projections: $m_s$ = $\ket{0}$ and $\ket{\pm1}$. Absorption of a photon preserves the spin state, but upon decay back to the ground state, radiative emission is more probable from the $\ket{0}$ state than from the $\ket{\pm1}$ states. This is because the $\ket{\pm1}$ states preferentially decay through intersystem crossing to a metastable singlet state. The resulting imbalance in radiative pathways enables optical detection of the spin state via differences in fluorescence intensity. Transitions between spin states can be driven using resonant microwave fields, where a transition from $\ket{0}$ to either $\ket{-1}$ or $\ket{+1}$ reduces the average fluorescence — a technique known as optically detected magnetic resonance (ODMR).The system’s suitability for magnetometry arises from the electron spin Hamiltonian (neglecting hyperfine interactions):

\begin{equation}
\label{nv_hamiltonian}
    H / \hbar = DS_{z}^{2} + \hbar \gamma \textbf{S}\cdot\textbf{B}.
\end{equation}

Here, $D$ represents the zero-field splitting, originating from the dipolar interaction between the two unpaired electrons, which is 2.877 GHz in the low-temperature, zero-phonon limit\cite{ivady_temperature_pressure}. The term $\gamma$ is the electron gyromagnetic ratio of approximately 28 GHz/T, describing the Zeeman interaction between the electron spin and an external magnetic field $B$. The Zeeman interaction lifts the degeneracy between the $\ket{-1}$ and $\ket{+1}$ states (Figure \ref{Figure3}c), with the resulting energy splitting providing a direct measure of the magnetic field component along the NV axis\cite{wolf_subpicotesla_2015}. See also Supplementary section S2 for more information about the spin state's splitting with magnetic field. In our experiment, a diamond sample (ThorLabs DNVB-1, 300 ppb NV concentration, size (l $\times$ w $\times$ h): 3 mm $\times$ 3 mm $\times$ 0.5 mm) containing an ensemble of NV centers in four crystallographic orientations was used (Figure \ref{Figure3}b). The diamond is cut along the (100) plane, ensuring that an upward-directed magnetic field projects equally onto all four NV axes.\\
\\
For magnetic field measurements near the coil, we placed the diamond between the flat side of the flux concentrator and a microfabricated microwave stripline antenna (sputtered gold on chromium sticking layer). This transparent loop-on-glass antenna (see Supplementary Figure S5) allows simultaneous excitation of the NV centers with 532 nm laser light (turned to circularly polarized by $\lambda$/4 plate to excite all populations equally) and collection of their fluorescence through a microscope objective (10x, 0.25 NA) immersed in liquid helium in a cryostat (see Figure \ref{Figure3}a for the setup schematic). Because we can control the cryostat temperature, it is possible to operate in a regime where the niobium core remains in its normal state while the niobium-titanium coil wiring is already superconducting — owing to their differing critical temperatures.\\ 
\\
While tuning the coil current up to 500 mA, we recorded ODMR spectra near the surface of the microwave loop at 0.5 mm height from the flux concentrator core. The resulting heatmap of these series of ODMR spectra shows a linear Zeeman shift of 0.013 $\pm$ 0.001 mT/mA (Figure \ref{Figure3}d). This value agrees well with finite-element simulations of the core in its normal state ($\sim$0.011 mT/mA). For the normal state measurement, given the large size of the outer coil, the field is homogeneous in the sampling area which is determined by the microscope objective’s point spread function, offering an in-plane resolution of $\sim$10 $\mu$m. To improve depth discrimination and suppress inhomogeneous ODMR broadening at high fields, a 75 $\mu$m pinhole was inserted into the detection path — a standard confocal microscopy practice (see Supplementary Figure S7). We estimate the out-of-plane resolution to be in the order of $\sim$50 $\mu$m (see Supplementary Section 2).\\

\begin{figure}
    \centering
    \includegraphics[trim={0 6.8cm 0 0}, clip, width=0.92\textwidth]{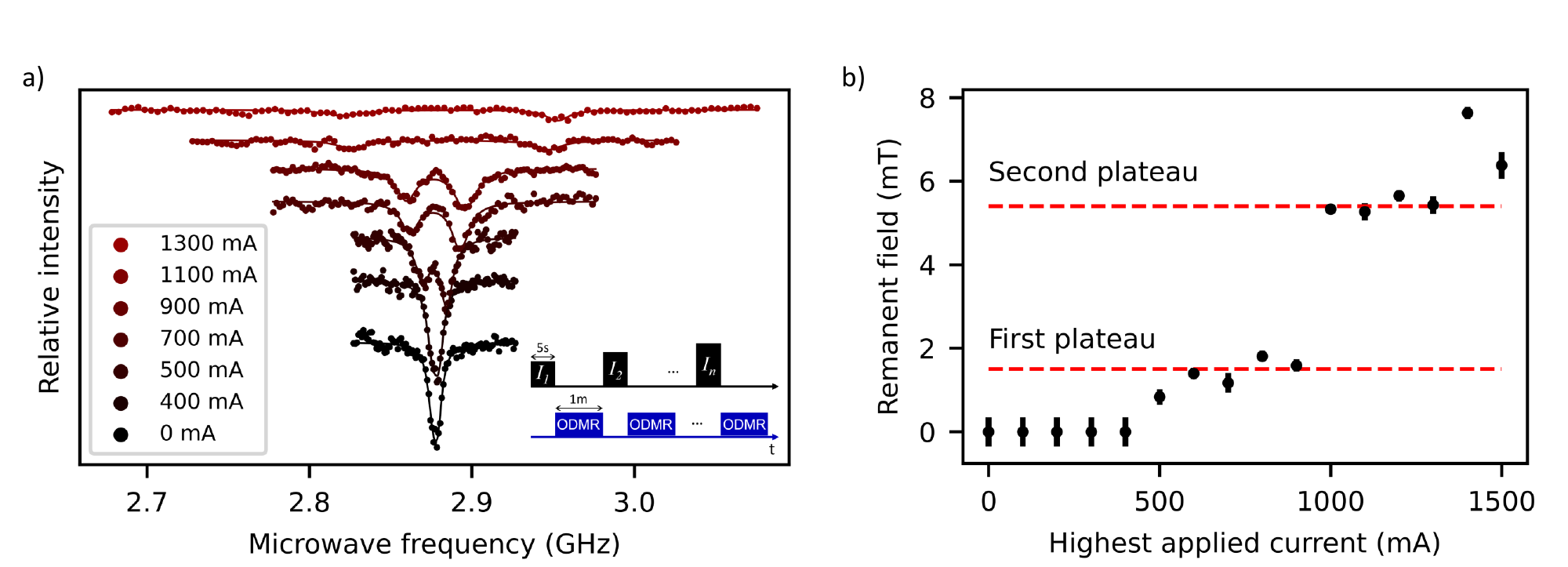}
    \caption{\textbf{Remanent fields in the flux concentrator after high-current operation}. \textbf{(a)} Six ODMR spectra taken after applying the labeled amount of current for 5 seconds (see also the inset measurement scheme). \textbf{(b)} The single spectra (including the subset in (a)), have been fitted to two Lorentzian curves, resulting in an estimate for the remanent magnetic field magnitude when solved for Equation \ref{nv_hamiltonian} The remanent field has been plotted as a function of the highest applied current to the flux concentrator.}
    \label{Figure4}
\end{figure}

Upon lowering the temperature below the critical temperature of niobium, the behavior of the magnetic field changes (Figure \ref{Figure3}e). The ODMR spectra reveal unequal splittings for the four NV orientations, indicating significant field curvature even at this distance above the slit, a result from the strong confinement of the magnetic field to the narrow slit region (see Supplementary Figure S8b). The measured Zeeman splitting for the superconducting core, as determined from the NV orientation that most aligns with the field, is 0.043 $\pm$ 0.005 mT/mA - about 3.3 $\pm$ 0.4 times higher than in the normal state. This enhancement slightly exceeds simulation predictions, likely due to the NV centers residing within a range of heights lower than the nominal 0.5 mm plane.\\   
\\
To investigate flux trapping, we applied incrementally increasing coil currents, holding each value for 5 seconds, then reducing the current to zero and finally recording an ODMR spectrum. This process was repeated in 100 mA steps up to 1.5 A. As shown in Figure \ref{Figure4}a, the ODMR spectrum does not measure a significant field up to 400 mA. However, above this current, a small remanent field has developed, which remained stable until $\sim$900 mA, beyond which it increased sharply again. At 1 A, a remanent field of $\sim$5.5 mT persisted at 0.5 mm above the core — comparable to the field generated during continuous operation at 130 mA. This remanent field likely results from flux trapping in the niobium core when the local magnetic field exceeds the lower critical field ($B_{c,1}$). Furthermore, the equivalent projection of the remanent field onto all NV axes reveals that there is little to no curvature in the field as opposed to the current-induced field in the flux concentrator. This could indicate that the remanent field is created by flux penetration and trapping deeper into the core. About a millimeter further outwards of the flux concentrator's center we measure a significantly weaker remanent field, indicating the remanent field mostly concentrates in the center of the loop, as expected from the localized breaching of the lower critical field. 
\\

\section{Discussion}

The flux trapping effect offers a plausible explanation for several phenomena observed in the levitation experiments. Notably, the levitating particle’s resonance frequency varied by up to 10\% between measurements performed with identical current parameters but following different histories of applied current. This is consistent with the percentage of remanent field from prior high-current operation. The presence of a remanent field could also explain why the observed mode frequencies in Figure \ref{Figure2}a are higher than expected from the simulations. Moreover, trapped flux is a known source of damping for levitating magnets or superconducting particles\cite{brandt_friction_1988, grosser_damping_2000, wang_dynamics_2019} and likely accounts for the sharp drop in quality factor succeeding high-current application. Strong remanent fields and the close proximity of the superconducting core to the levitating particle likely make flux trapping a prominent damping mechanism in this trap.  \\
\\
While simulations predicted that local fields should reach 350 mT at 1.7 A — with the lower critical field ($B_{c,1}$ = 173.5 mT\cite{stromberg_superconducting_1965}) being breached around 0.85 A — the first experimental signs of flux trapping appeared between 0.4 A and 0.5 A (first plateau in Figure \ref{Figure4}b). Although the calculated field at 0.5 A was $\sim$100 mT (assuming ideal smooth edges), real-world machining imperfections in the brittle niobium likely enhanced local field concentrations at rough edges, triggering earlier flux penetration. The pronounced increase in trapped flux observed between 0.9 A and 1 A may reflect a more widespread critical field breach around the entire loop, capacitated by its sharp edges (second plateau in Figure \ref{Figure4}b).\\
\\
To prevent premature flux trapping, future designs might employ a flux concentrator made from a more malleable material, later polished mechanically or chemically and coated with a high-purity superconducting layer. Since no noticeable flux trapping was detected up to 400 mA, this may define a safe operating range. Reducing the vertical separation between the coils could further increase the field gradient without elevating the local field at the sensitive core edges. For example, decreasing the separation from 1.2 mm to 200 $\mu$m for the same 50 $\mu$m particle would raise the vertical field gradient from 56 T/m to approximately 175 T/m per 1 A of current, allowing for 60 Hz per 1 A for the x-mode frequencies and $>$150 Hz for the z-mode frequencies (see Figure S10 in Supplementary). 
Another alternative is the use of a type I superconductor to avoid flux trapping, but the critical field that can be achieved is significantly lower at 80.3 mT for lead\cite{chanin_critical-field_1972}. Finally, we emphasize that on-chip flux concentrators could be an appealing alternative to planar on-chip coils, potentially with flip-chip designs\cite{paradkar_superconducting_2025}. Such a design would also allow significant downscaling of the flux concentration region, resulting in increased magnetic field gradients and thus stiffer trapping potentials with vibrational eigenfrequencies into the kilohertz regime. Moreover, the ability of depositing high-purity thin films of superconductors could reduce available defect sites for flux trapping. 

\section*{Conclusion}

We have demonstrated the design and operation of a superconducting flux concentrator trap capable of levitating superconducting microparticles at millikelvin temperatures. However, during levitation experiments, we observed shifts in the particle’s vibrational resonance frequencies and a marked reduction in quality factors following the application of higher coil currents. These effects suggested the presence of persistent, remanent magnetic fields within the trap, likely caused by flux trapping in the superconducting core once local fields exceeded the lower critical field threshold.\\
\\
With NV center magnetometry we confirmed the onset of flux trapping above 500 mA of applied current, with remanent fields persisting after current removal and increasing sharply around 0.9–1.0 A. These observations correlated with both the shifts in levitating particle dynamics and the significant drops in quality factors.\\ 
\\
Based on these results, we propose that future flux concentrator designs should consider alternative materials, improved surface treatments or on-chip designs to mitigate premature flux trapping. This work highlights both the capabilities and limitations of superconducting flux concentrator traps and illustrates the power of NV-based magnetometry for diagnosing and optimizing complex cryogenic magnetic systems.

\section*{Acknowledgements}
We thank Merlijn Camp for the fabrication of the flux concentrator, Çağan Karaca for prior measurements on flux concentrators and Michel Orrit for providing access to his laser and cryostat. This work was supported by the European Union (ERC StG, CLOSEtoQG, Project 101041115).

\section*{Author Contributions}
RS, MJ, EvdB, TdJ, KH, JP, TO and BH conceptualized, and RS, MJ, TdJ and BH built the experimental set-up for the flux concentrator trap. RS, MJ, EvdB, TdJ and BH conducted the levitation measurements. RS and EvdB analyzed the levitation data. RS, MJ, EvdB and BH performed numerical simulations of the trap. RS, MJ and BH conceptualized, and RS and MJ built the experimental set-up for the NV center magnetometry. RS and MJ conducted the NV magnetometry measurements and analyzed the data. RS prepared the figures and wrote the manuscript with input from all authors. BH supervised the project.

\section*{Competing interests}
The authors declare no competing interests.

\section*{Additional information}
\textbf{Supplementary information} The online version contains supplementary material.\\

\textbf{Correspondence} Correspondence and requests for materials should be addressed to Bas Hensen.\\

\textbf{Data availability} The data that support the findings of this study are available from the corresponding author upon reasonable request.\\

\section*{References}
\bibliography{refs}

\end{document}


\title{Superconducting flux concentrator coils for levitation of particles in the Meissner state: supplementary material}

\date{June 2025}

\author{Robert Smit}
\author{Martijn Janse}
\author{Eli van der Bent}
\author{Thijmen de Jong}
\author{Kier Heeck}
\author{Jaimy Plugge}
\author{Tjerk Oosterkamp} 
\author{Bas Hensen} \email{hensen@physics.leidenuniv.nl}
\address{Leiden Institute of Physics, Leiden University, P.O. Box 9504, 2300 RA Leiden, The Netherlands}

\maketitle

\section{Magnetic levitation}\label{SM_A}

\begin{figure}
    \centering
    \includegraphics[scale=0.45]{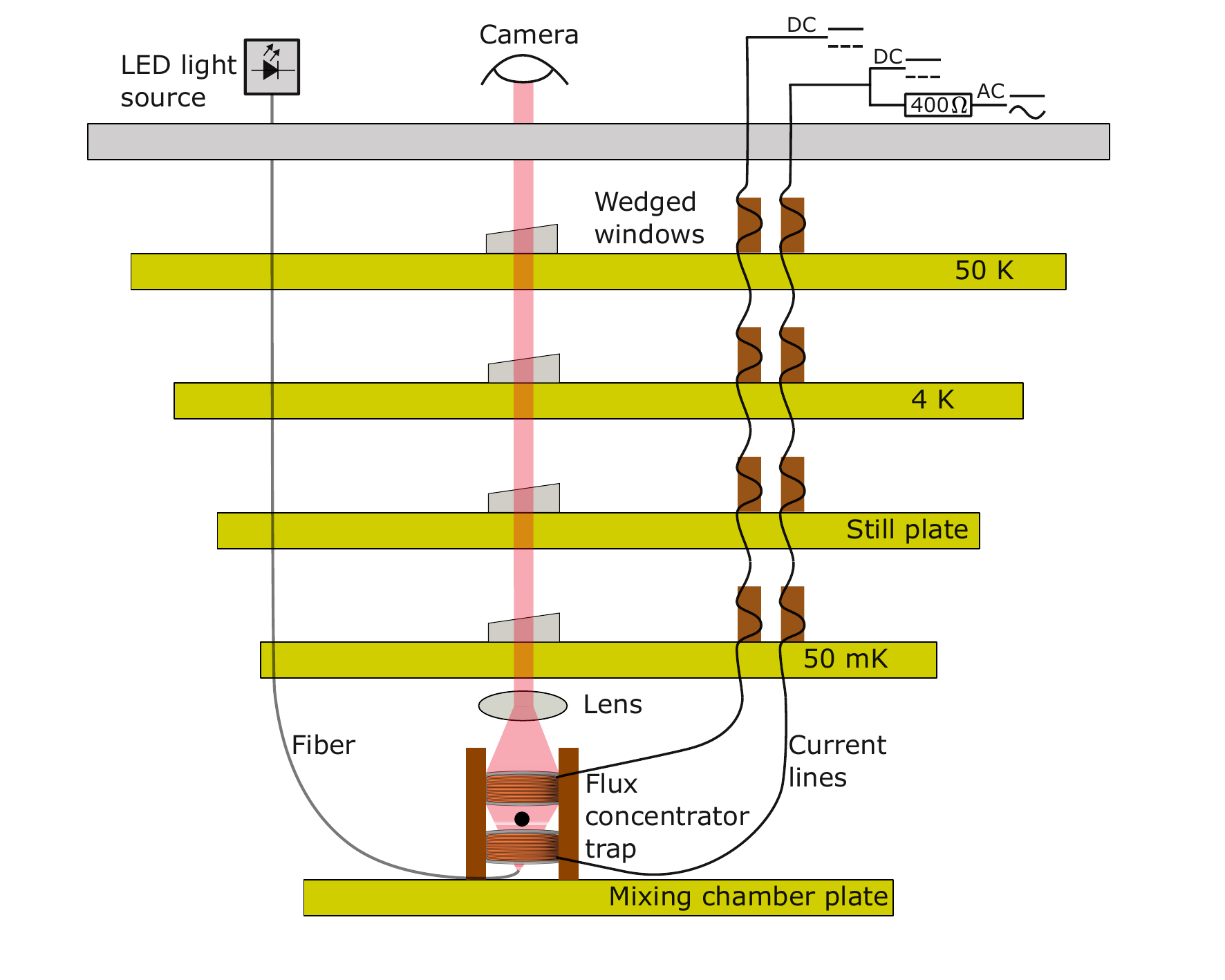}
    \caption{Schematic of the dilution refrigerator setup with an optical viewport used for the levitation experiments.}
    \label{Figure_mK_setup}
\end{figure}

The levitation experiments were carried out in a dilution refrigerator from Leiden Cryogenics (model LC CF-CS110), which achieves a standard base temperature of approximately 10 mK. A schematic of the experimental setup within the cryostat is shown in Figure \ref{Figure_mK_setup}. When the maximum illumination intensity from a red LED is coupled into the optical fiber (around 35 $\mu$W), the temperature rises modestly and stabilizes between 30 and 35 mK.\\
\\
The currents for the levitation coils are supplied by two programmable Tenma power supplies. No additional filtering or current stabilization was implemented in our setup, so the motional resonance frequencies are likely broadened by spectral diffusion caused by current noise. To mitigate such fluctuations and achieve higher quality factors, Hofer \textit{et al.} employed a low-pass filter with a sub-Hz cutoff frequency alongside feedback control on the current source output\cite{hofer_high-q_2023}. In our case, for the lower quality factors measured, current noise is likely not the primary limiting factor, as higher $Q$ values have been observed in standard superconducting coil levitation experiments that we performed under similar conditions.\\

Before cooling down the cryostat, we position a lens above the flux concentrator trap to ensure proper focus. Light is collected through a series of fused-silica windows mounted on each temperature stage of the cryostat. These windows block thermal radiation while transmitting visible light and are wedged to prevent interference effects such as etaloning. The collected light is then focused onto a CS505MU camera, with image acquisition performed using ThorCam software or through a python script. By optimizing the acquisition parameters, frame rates of up to 300 fps are achieved — more than sufficient for resolving the slit mode.\\

\begin{figure}
    \centering
    \includegraphics[scale=0.4]{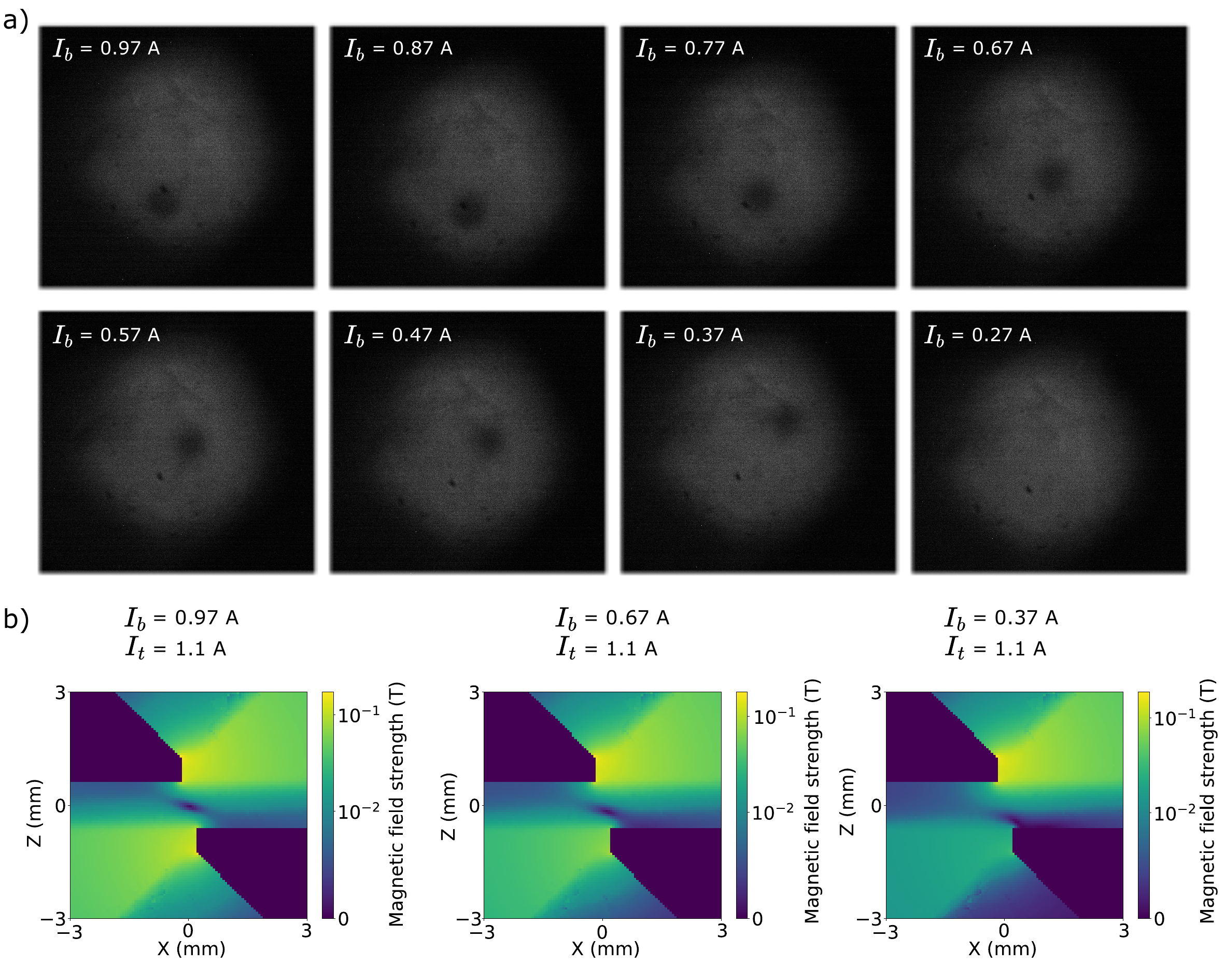}
    \caption{The top part shows eight captures of the trap center while a fixed top coil current of 1.1 A and a variable bottom coil current, shown as $I_b$, are applied. The particle's position clearly shifts along the slit direction due to the asymmetry between the two coil currents. At the lowest bottom coil current of 0.27 A the particle is not trapped anymore. Corresponding simulations of the shifting potential minimum for three of those bottom coil currents are shown at the bottom.}
    \label{Figure_particle_position}
\end{figure}

\begin{figure}
    \centering
    \includegraphics[scale=0.4]{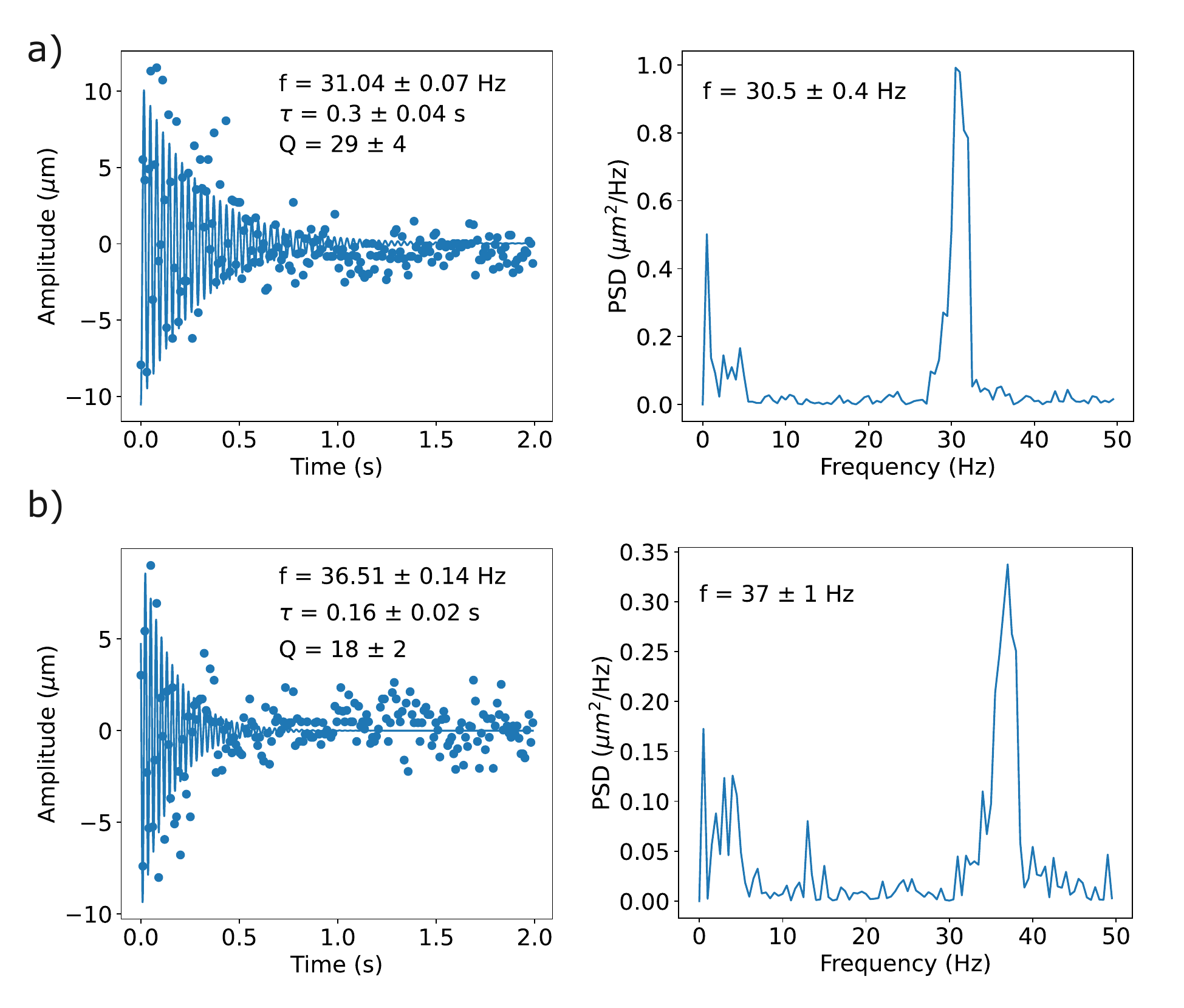}
    \caption{Ringdown measurements and power spectral density curves obtained with video analysis for different currents. The currents for the bottom and top coil were respectively 1.2 A and 1.6 A in (a) and 1.4 A and 1.7 A in (b).}
    \label{Figure_ringdowns}
\end{figure}

Examples of recorded trap images are presented in Figure \ref{Figure_particle_position}. In these image sequences, the current in the top coil is held constant while the current in the bottom coil is varied to increase the asymmetry between the two. As shown by the simulations in the lower part of Figure \ref{Figure_particle_position}, due to the shape of the trapping potential, the particle’s motion involves both in-plane displacement (well resolved in the top-view imaging) and out-of-plane movement (which is less discernible due to limited depth resolution).\\
\\
For dynamic measurements such as ringdowns, we optimize video acquisition by first driving the particle at its resonance frequency for a set period, then ceasing the drive while continuing to record the motion during the ringdown phase. Each video frame is analyzed by fitting the particle’s center position, providing a time series of its motion. This data reveals clear oscillatory behavior and a subsequent decay in amplitude. We fit these time series to an exponentially decaying sinusoid, characterized by a resonance frequency $f_0$ and a decay time constant $\tau$. The mechanical quality factor is then calculated using $Q = \pi f_0 \tau$. Two representative examples of these measurements are shown in Figure \ref{Figure_ringdowns}.\\

\section{NV center magnetometry}

\subsection{Theory}

The nitrogen-vacancy (NV) center spin in diamond is an effective probe for detecting magnetic fields, owing to its efficient optical readout via fluorescence. Transitions between the electron spin states $\ket{0}$ and $\ket{\pm1}$ can be observed through optically detected magnetic resonance (ODMR), appearing as dips in the fluorescence signal with a contrast of up to 30\%. In the absence of a magnetic field, both electron spin resonances occur at 2.877 GHz (at 0 K) and are degenerate. When a magnetic field is applied along the NV center axis, the $\ket{-1}$ state shifts to lower frequencies, while the $\ket{+1}$ state shifts to higher frequencies. The resulting splitting between these levels serves as a direct measure of the local magnetic field.\\
\\
Magnetic field sensing can be performed using a single NV center or an ensemble within a diamond crystal. In a single-crystal diamond, NV centers naturally occur along four crystallographic orientations, determined by the diamond lattice structure. By analyzing the magnetic response of the NV centers along each of these four orientations, it is possible to reconstruct the full magnetic field vector. If the NV axis is taken as the principal $z$  for each orientation, the Hamiltonian describing the system is given by:

\begin{equation}
H / \hbar = DS_{z}^{2} + \gamma \textbf{B} \cdot \textbf{S}.
\end{equation}

Here, $D$ is the zero-field splitting of 2.877 GHz, $\gamma$ the gyromagnetic ratio and $S_z$ and $\textbf{S} = (S_x, S_y, S_z)$ are Pauli matrices: 

\begin{equation}
S_x = \frac{1}{\sqrt{2}} 
\begin{pmatrix}
0 & 1 & 0 \\
1 & 0 & 1 \\
0 & 1 & 0
\end{pmatrix} , 
S_y = \frac{1}{\sqrt{2}}
\begin{pmatrix}
0 & -i & 0 \\
i & 0 & -i \\
0 & i & 0
\end{pmatrix},
S_z = 
\begin{pmatrix}
1 & 0 & 0 \\
0 & 0 & 0 \\
0 & 0 & -1
\end{pmatrix}.
\end{equation}

To determine the local magnetic field, this Hamiltonian must first be diagonalized to obtain the energy eigenvalues of the three spin states. Since the magnetic field can be characterized by an angle $\theta$ relative to the NV axis, all combinations of field components in the $x$ and $y$
directions that produce the same angle form a degenerate cone in space. For simplicity, the magnetic field vector can be expressed as: 

\begin{equation}
    \textbf{B} = B_0 (sin\theta \hat{x} + cos\theta \hat{z}).
\end{equation}

\begin{figure}
    \centering
    \includegraphics[scale=0.35]{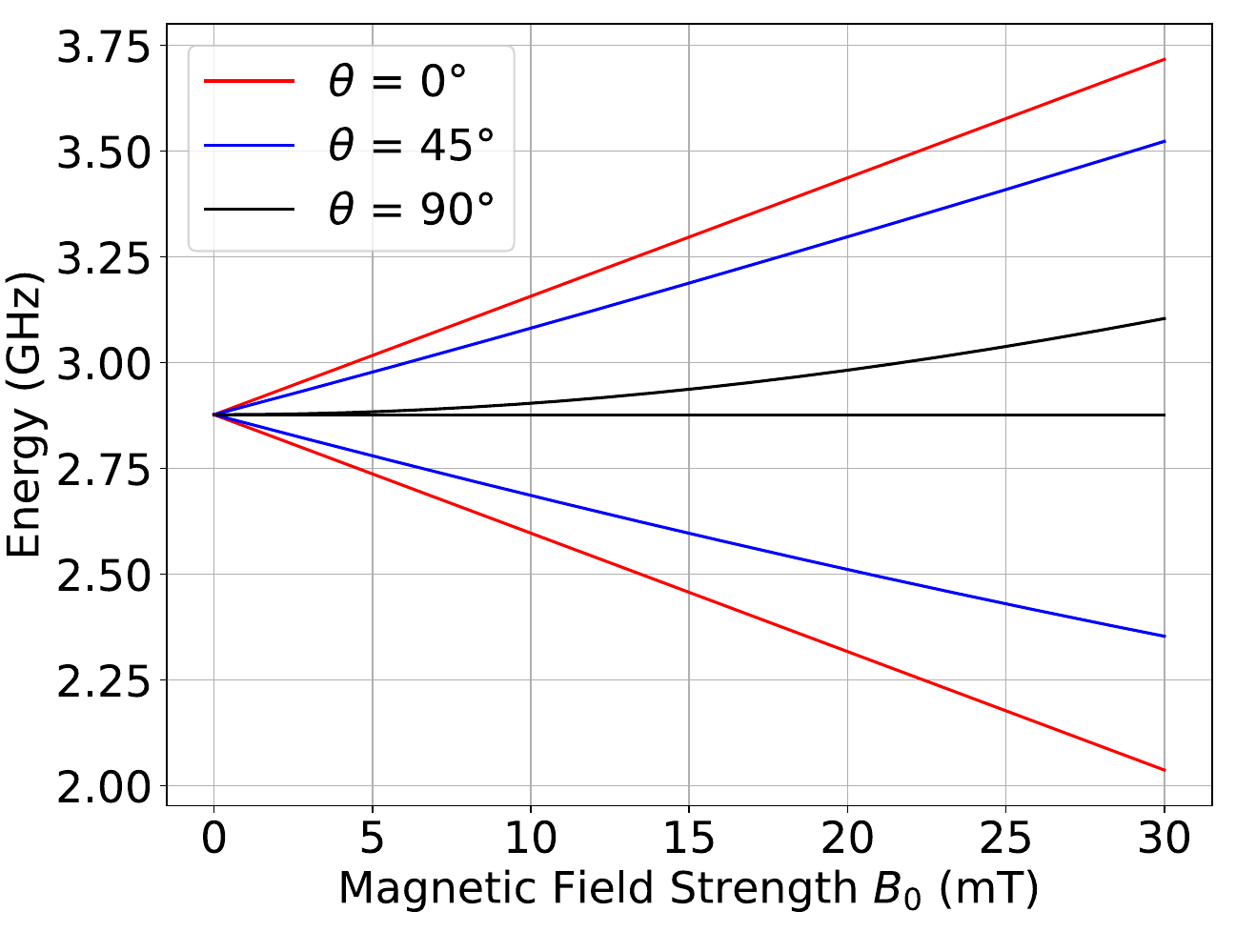}
    \caption{Energy eigenvalues for the $\ket{-1}$ and $\ket{+1}$ states as a function of magnetic field strength, calculated by solving equation 1 with angular dependence of the field given by equation 3. The three colors represent values for which the magnetic field is along the NV axis (red line), orthogonal to the NV axis (black line) and at an in-between angle (blue line).}
    \label{Figure_NV_eigenvalues}
\end{figure}

When the magnetic field is aligned with the NV axis ($\theta$ = 0), the splitting of the spin levels is linear, as illustrated by the red line in Figure \ref{Figure_NV_eigenvalues}. As the angle increases, the splitting becomes nonlinear, and asymmetries relative to the zero-field splitting become more pronounced at higher magnetic fields (as shown in Figure \ref{Figure_ODMR_sweep}). Furthermore, due to spin state mixing at elevated fields, a measurable splitting persists even when the field is perpendicular to the NV axis.\\
\\
In measurements involving the flux concentrator’s magnetic field, contributions from all four NV orientations must be considered, each with its distinct resonance response as shown in Figure \ref{Figure_NV_eigenvalues}. The diamond used in this experiment is cut along the (100) crystallographic plane, making the in-plane and out-of-plane directions symmetry axes. In this configuration, the out-of-plane direction forms an angle of approximately 54.7$^{\circ}$ with each NV axis, while the in-plane direction is at an angle of 35.3$^{\circ}$. This geometry ensures that for out-of-plane magnetic fields — which is typically the case in our setup — the field components project equally onto all four NV orientations.

\subsection{Experiments}

\begin{figure}
    \centering
    \includegraphics[scale=0.3]{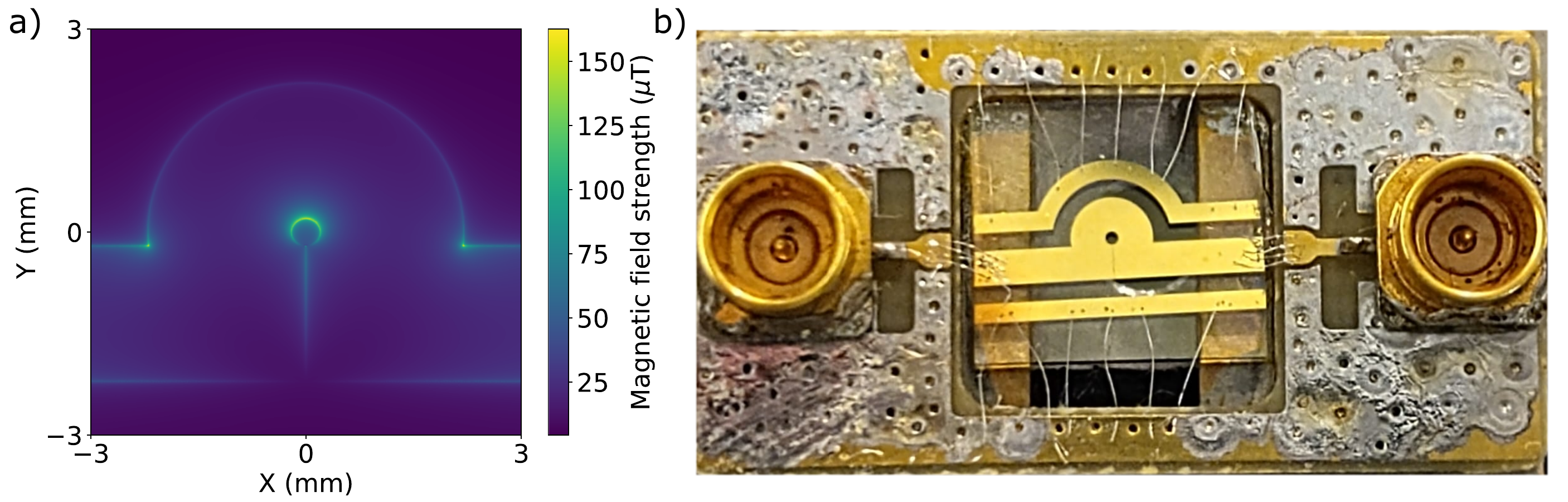}
    \caption{Panel (a) shows a simulation of the RF magnetic field $B_1$ in the plane of the microwave loop when the maximum expected current at 25 dB of RF power runs through the loop. The field in the center corresponds to about 50 $\mu$T and would give rise to at most 1.4 MHz Rabi flopping rate $\Omega = \gamma B_1$. The corresponding microwave loop used in the experiments is shown in panel (b).}
    \label{Figure_microwave_loop}
\end{figure}

A schematic of the NV magnetometry setup is shown in Figure 3 of the main text. The system is built around a confocal fluorescence microscope integrated with a liquid-helium cryostat (Janis SVT), capable of reaching base temperatures of approximately 1.4 K. To drive spin transitions, we fabricated a microwave coplanar waveguide featuring a 400 $\mu$m diameter loop on single-side ITO-coated glass (see Figure \ref{Figure_microwave_loop}), patterned using electron-beam lithography. To prevent charging during the patterning, a 5 nm chromium layer was deposited on the non-conductive side of the glass, which was later removed via a chromium etch. The microwave structures, consisting of 5 nm chromium / 100 nm gold, remained intact after etching.\\
\\
The processed glass was mounted on a sample holder and connected to a PCB using wire bonds. This was further interfaced with an RF source (R\&S SGS100a) capable of delivering up to 25 dB of RF power. The microwave loop was designed for a characteristic impedance of 50 $\Omega$, with measured transmission losses ($S_{21}$) of only 2–4 dB across a 2–4 GHz frequency range. To achieve optimal ODMR contrast, which reached a maximum of 10\%, the microwave signal was pulsed for a few microseconds with a duty cycle of 10\%. This approach reduced heat dissipation in the microwave loop, preventing helium boiling and minimizing fluorescence signal fluctuations.\\

\begin{figure}
    \centering
    \includegraphics[scale=0.3]{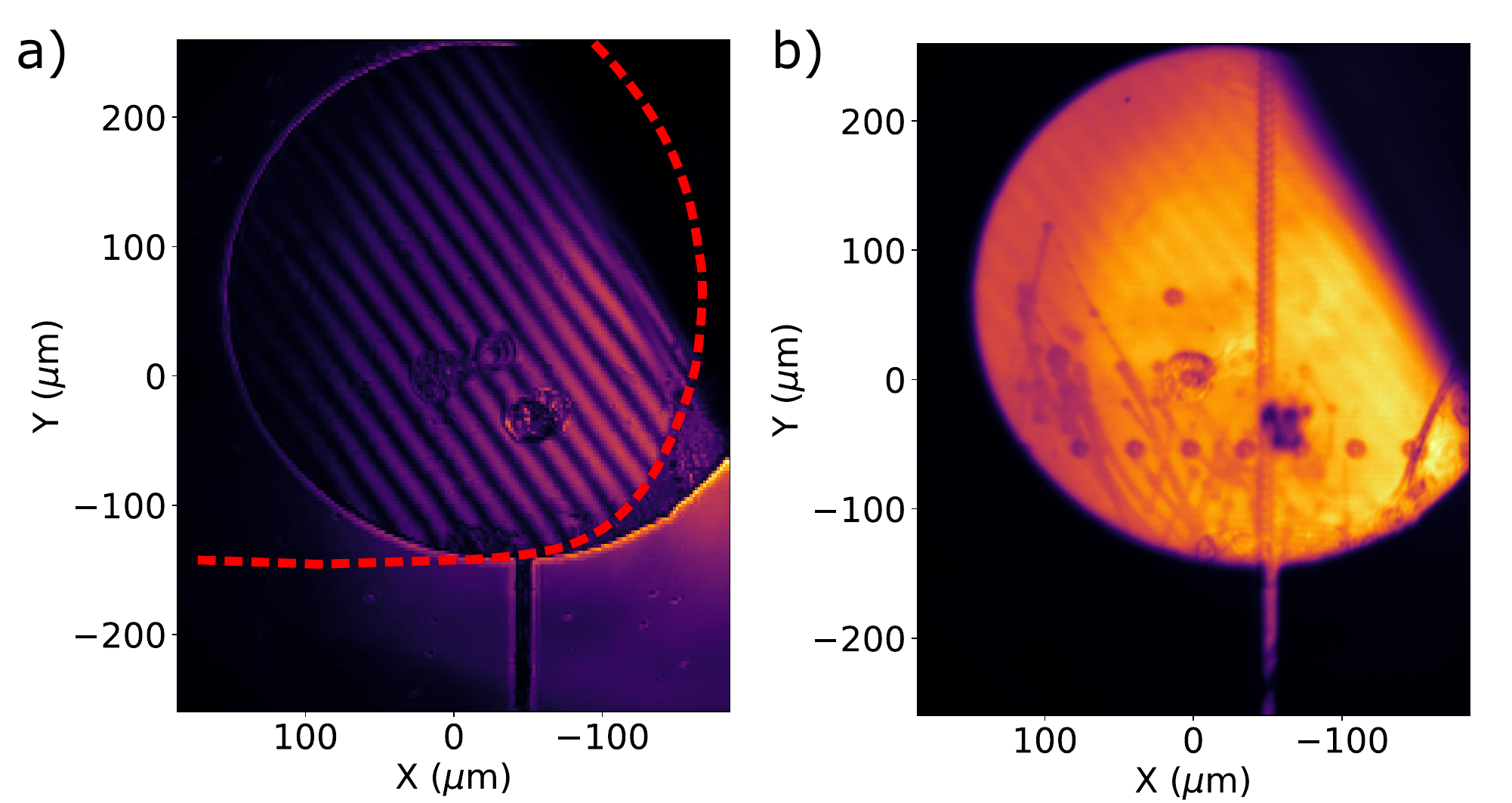}
    \caption{Confocal scans of reflected light in (a) and fluorescence in (b). The orientation of the inner loop of the flux concentrator is outlined with the dashed line. The slit of the flux concentrator is oriented to the left side of the image. The darker region in the upper right corner is due to a break in the glass of the microwave loop during cooldown.}
    \label{Figure_confocal_scans}
\end{figure}

\begin{figure}
    \centering
    \includegraphics[scale=0.3]{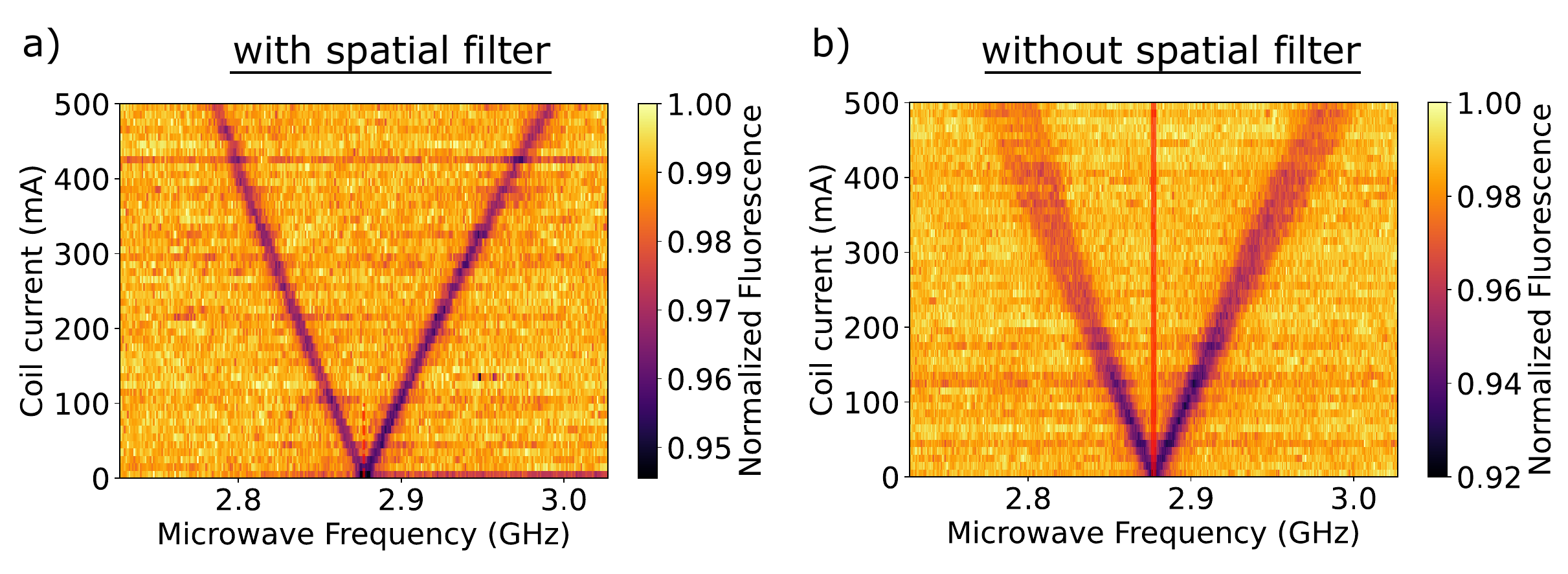}
    \caption{Series of ODMR spectra for a range of currents through the coil. The spectrum in (a) was taken with a pinhole in the detection path, while the spectrum in (b) was taken without.}
    \label{Figure_ODMR_sweep}
\end{figure}

The diamond was imaged through the microwave loop, so fluorescence was collected only from the region within the loop (see Figure \ref{Figure_confocal_scans}b). An objective lens (10x, 0.25 NA from Melles Griot) immersed in liquid helium collected the emitted fluorescence, which was then directed through a dichroic mirror and additional optical filters before detection by a single-photon counting module (SPCM-AQRH-14). The detected signal was subsequently processed by an Adwin Gold control system.\\ 
\\
The spatial resolution of the magnetometry measurements was primarily limited by the optical resolution of the microscope, which depends on both the numerical aperture (NA) of the objective and the system’s depth of field. To improve the depth of field and mitigate spectral broadening caused by magnetic field gradients, a pinhole (75 $\mu$m) was inserted into the detection path. The resulting improvement is clearly illustrated in Figure \ref{Figure_ODMR_sweep}, which compares data acquired with and without the pinhole.\\
\\
Although the exact lateral resolution of the laser spot wasn’t directly measured, an upper estimate can be inferred from Figure \ref{Figure_confocal_scans}b. In this image, a vertical line and a series of spots along a horizontal line indicate positions where prolonged measurement led to photobleaching. By examining one of these isolated bleached spots, a fluorescence bleaching profile with a diameter of approximately 10 $\mu$m was determined — about four times larger than the diffraction limit for a 0.25 NA objective at 532 nm, which corresponds to a 2.6 $\mu$m beam waist.\\

\begin{figure}
    \centering
    \includegraphics[scale=0.4]{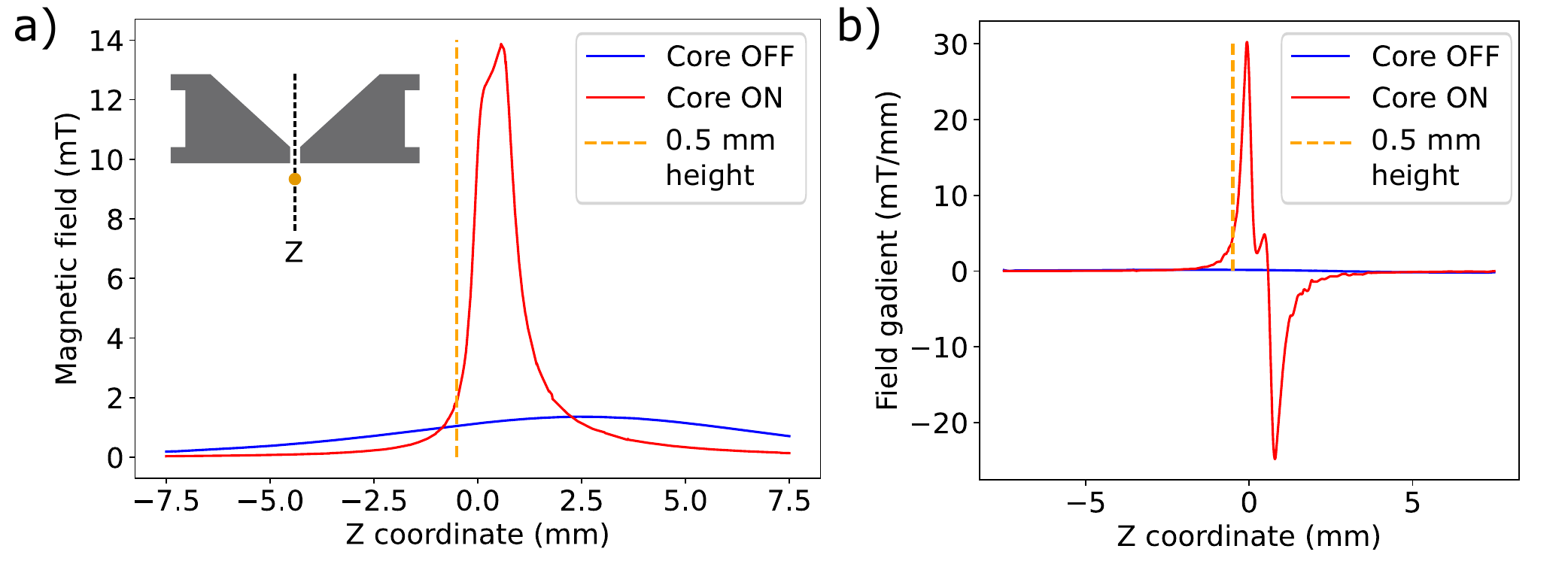}
    \caption{Panel (a) shows the magnetic field strength in the $z$ direction for a superconducting and non-superconducting core. The inset illustrates how the $z$ coordinate is defined. The ratio of the two maxima is close to 10 - the amplification factor. The position of the maximum in $z$ is different. Panel (b) shows the magnetic field gradient as a function of the $z$ coordinate. The orange dashed line indicates the height at which the magnetic field measurements were taken.}
    \label{Figure_field_gradients}
\end{figure}

Estimating the axial resolution is more challenging, but theoretically it can be expressed as: $\Delta z = 2n\lambda / NA^2$. Given the high refractive index of diamond (n = 2.4), the axial resolution for the 0.25 NA objective is at least 41 $\mu$m. This limited axial resolution makes measurements susceptible to magnetic field gradients, which were calculated for the flux concentrator in Figure \ref{Figure_field_gradients}, both with and without a superconducting core, at a driving current of 100 mA. At the measurement height of 0.5 mm above the flux concentrator loop (corresponding to $z$ = -0.5 mm in Figure \ref{Figure_field_gradients}), the magnetic field gradient was approximately 0.166 mT/mm with the core in the normal state, increasing to 4.45 mT/mm in the superconducting state.\\
\\
The corresponding ODMR spectra reflected these gradients: with a normal-state core (weak gradient), the observed linewidth was about 8 $\pm$ 2 MHz, while with a superconducting core at 100 mA, the linewidth broadened to 28 $\pm$ 4 MHz. The measured field strength of 3 $\pm$ 0.3 mT, combined with the observed linewidth increase, indicated a local field gradient of approximately 0.22 $\pm$ 0.02 mT. From this, the effective axial resolution was estimated to be around 50 $\mu$m — slightly larger than the theoretical limit but consistent with the gradient-induced broadening effects.

\section{Optimization of the flux concentrator trap}

\begin{figure}
    \centering
    \includegraphics[scale=0.3]{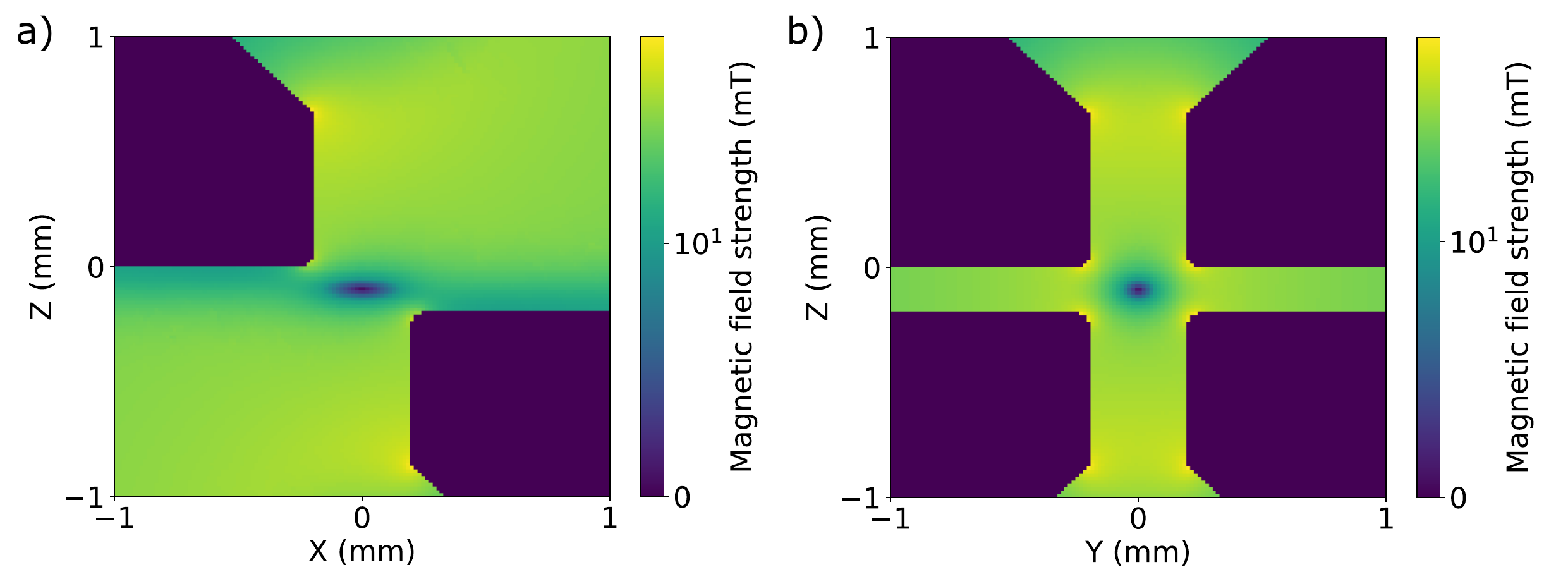}
    \caption{Simulations of the XZ plane (a) and YZ plane (b) of the anti-Helmholtz field for a configuration in which the radius of the flux concentrator loop matches the vertical separation between the coils.}
    \label{Figure_optimizations}
\end{figure}

\begin{figure}
    \centering
    \includegraphics[scale=1]{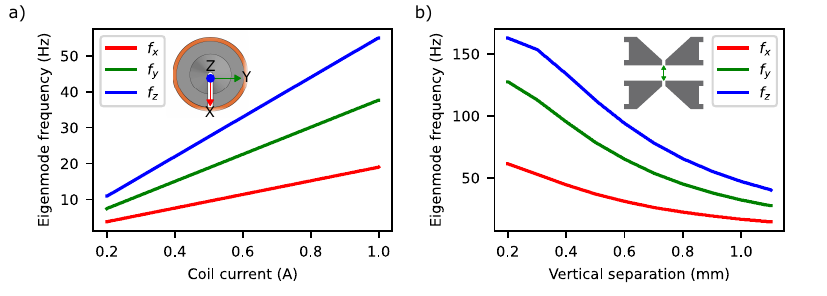}
    \caption{Simulated eigenfrequencies for the three eigenmodes as a function of coil current in (a) and as a function of vertical separation between the coils in (b). The values in (b) correspond to a simulated current of 1 A through both coils.}
    \label{Figure_simulation_eigenfreq}
\end{figure}

With simulations of the flux concentrator trap, we find that the optimal symmetry of the potential minimum is reached when the radius of the inner loop of the flux concentrator matches the spacing between the individual flux concentrators (Figure \ref{Figure_optimizations}). The further decrease of the vertical separation between the flux concentrators, while keeping currents fixed, increases the eigenmode frequencies, as shown in Figure \ref{Figure_simulation_eigenfreq}b. The increase of the eigenmode frequencies scales non-linearly with vertical separation and indicates that also the geometric factors of the trap are increasing and thus become more close to those found for an ideal anti-Helmholtz configuration\cite{hofer_analytic_2019}.

\section*{REFERENCES}
\bibliography{refs}